\documentclass[manuscript]{acmart}
\AtBeginDocument{%
  \providecommand\BibTeX{{%
    \normalfont B\kern-0.5em{\scshape i\kern-0.25em b}\kern-0.8em\TeX}}}

\setcopyright{acmcopyright}
\copyrightyear{2025}
\acmYear{2025}
\acmDOI{XXXXXXX.XXXXXXX}

\acmConference[XXX]{XXX}{XXX}{XXX}
\acmBooktitle{XXX} 
\acmPrice{15.00}
\acmISBN{978-1-4503-XXXX-X/18/06}

\usepackage{bm} 

\begin{document}


\title{To Trust or Distrust AI: A Questionnaire Validation Study} 



\author{Nicolas Scharowski}
\authornote{Both authors contributed equally to this research.}
\authornote{Corresponding Author.}
\email{nicolas.scharowski@unibas.ch}
\orcid{0000-0001-5983-346X}
\affiliation{
  \institution{Center for General Psychology and Methodology, University of Basel}
  \streetaddress{Missionsstrasse 62a}
  \city{Basel}
  \postcode{CH-4055}
  \country{Switzerland}
}

\author{Sebastian A. C. Perrig}
\authornotemark[1]
\email{sebastian.perrig@unibas.ch}
\orcid{0000-0002-4301-8206}
\affiliation{
  \institution{Center for General Psychology and Methodology, University of Basel}
  \city{Basel}
  \country{Switzerland}
}

\author{Lena Fanya Aeschbach}
\email{lena.aeschbach@unibas.ch}
\orcid{0000-0001-9092-6103}
\affiliation{
  \institution{Center for General Psychology and Methodology, University of Basel}
  \city{Basel}
  \country{Switzerland}
}

\author{Nick von Felten}
\email{nick.vonfelten@unibas.ch}
\orcid{0000-0003-0278-9896}
\affiliation{
  \institution{Center for General Psychology and Methodology, University of Basel}
  \city{Basel}
  \country{Switzerland}
}

\author{Klaus Opwis}
\email{klaus.opwis@unibas.ch}
\orcid{0000-0003-0509-8070}
\affiliation{
  \institution{Center for General Psychology and Methodology, University of Basel}
  \city{Basel}
  \country{Switzerland}
}

\author{Philipp Wintersberger}
\email{philipp.wintersberger@fh-hagenberg.at}
\orcid{0000-0001-9287-3770}
\affiliation{
  \institution{Univ. of Applied Sciences Hagenberg}
  \city{Hagenberg}
  \country{Austria}
}

\author{Florian Brühlmann}
\email{florian.bruehlmann@unibas.ch}
\orcid{0000-0001-8945-3273}
\affiliation{
  \institution{Center for General Psychology and Methodology, University of Basel}
  \city{Basel}
  \country{Switzerland}
}

\renewcommand{\shortauthors}{Scharowski and Perrig, et al.}


\begin{abstract}

Despite the importance of trust in human-AI interactions, researchers must often rely on questionnaires adapted from other fields, which lack validation in the AI context. Motivated by the need for reliable and valid measures, we investigated the psychometric quality of the most commonly used trust questionnaire in the context of AI by Jian, Bisantz, and Drury (2000). In a pre-registered online experiment (N = 1485), participants observed interactions with both trustworthy and untrustworthy AI and rated them. Our findings revealed that the data did not support the originally proposed single-factor structure for the questionnaire. Instead, results suggested a two-factor solution, distinguishing between trust and distrust. Based on our findings, we provide recommendations for future studies on how to use the questionnaire. Finally, we present arguments for considering trust and distrust as two distinct constructs, emphasizing the opportunities and added value of measuring both in human-AI interactions.

\end{abstract}

\ccsdesc[500]{Human-centered computing~Empirical studies in HCI}
\ccsdesc[500]{Human-centered computing~User studies}

\keywords{AI, Trust, Distrust, Measurement, Questionnaires, Validation, Psychometrics}


\maketitle

\section{Introduction}

With artificial intelligence (AI) becoming increasingly integrated into people's daily lives, there is a growing need for a comprehensive understanding and appropriate measurement of human trust in AI. Trust is not only an essential element in human-AI interactions as it shapes how people use and rely on AI \citep{hoff2015trust, lee2004trust}, but also a key motivation for research into explainable AI (XAI) to create more transparent AI systems \citep{lipton2018mythos}. Despite this importance, the operationalization and measurement of trust are complicated by various challenges.

For one thing, a multitude of different definitions and conceptualizations of trust exist \citep{benk2022value, vereschak2021evaluate, ueno2022scoping,muir1994trust} that are often not clearly distinguished from related terms (e.g., "reliance" \cite{poursabzi2018manipulating}, "situational trust" \citep{hoff2015trust}, "perceived trustworthiness" \citep{weitz2021let}, "calibrated trust" \cite{langer2021we} or "warranted trust" \cite{jacovi2021formalizing,hoffman2009dynamics}). Not clearly distinguishing between these terms can lead to theoretical entanglements and a divergent operationalization of trust \citep{kohn2021measurement}. 
For example, trust, viewed as an attitude \citep{lee2004trust}, is a subjective psychological construct, typically measured via questionnaires, also called survey scales \citep{scharowski2022trust}. Meanwhile, reliance, as a behavior \citep{lee2004trust}, can be assessed using more objective observational methods such as analyzing changes in an individual's behavior after being presented with an AI recommendation (e.g., switch ratios \citep{Lu2021, yin2019understanding}). 
Conceptualizations such as "calibrated" or "warranted" trust also emphasize that the motivation of XAI should not be to merely increase trust arbitrarily and unjustifiably. Instead, trust should be aligned and calibrated to the AI's trustworthiness \citep{lee2004trust, wischnewski2023measuring}. In this regard, trust is warranted when the AI is trustworthy and unwarranted when it is untrustworthy \citep{jacovi2021formalizing}. Although the importance of calibrated trust has been recognized by the community \citep{wischnewski2023measuring}, the corresponding perspective - that distrust in untrustworthy AI is also warranted – remains relatively underemphasized, despite being an integral factor motivating XAI \citep{jacovi2021formalizing}. Indeed, distrust seems a comparatively overlooked construct in current human-AI research \citep{ueno2022scoping}.

Beyond these theoretical challenges, empirical studies measuring trust often use single-items \citep[e.g.,][]{yu2017user} or develop their own questionnaires \citep[e.g.,][]{yin2019understanding, merritt2011affective}. However, self-developed questionnaires and single-items usually lack a rigid construction and quality assurance process and are often only used in an individual study, complicating comparing different study results \citep{furr2011scale}. Thus, it has been recommended to use validated trust questionnaires \citep{wischnewski2023measuring} whose psychometric quality (i.e., objectivity, reliability, and validity) has been scrutinized. 
But even if researchers address these challenges and use standardized questionnaires for measuring trust, they have to resort to and adapt scales from other disciplines, as there is no validated questionnaire for trust in AI. 
For example, it is common practice among researchers to use the \emph{Trust between People and Automation} scale (TPA) by \citet{jian2000foundations} and rephrase the questionnaires' items to fit the study context \citep{vereschak2021evaluate}. However, such practices raise concerns about whether the modified scale still measures what it was initially intended to measure \citep{furr2011scale, juniper2009modified}. 
In fact, most studies measuring human trust in AI do not report the psychometric quality of the questionnaires they used \citep{vereschak2021evaluate} and only recently, \citet{Lai_2023} pointed out that the research community lacks practices to validate and reuse standardized measurements. At best, this makes it challenging for other researchers to replicate or build upon existing work. At worst, using non-validated trust questionnaires in the context of AI can generate research results that do not withstand psychometric scrutiny and thus lead to ambiguous or inconsistent findings, impeding progress in XAI research.
Despite this need for standardized measures, the psychometric quality of the TPA remains to be thoroughly investigated in an AI context. Our research aims to fill this research gap by validating the questionnaire in a pre-registered online experiment, following current best practices for investigating scale quality.

The contribution of this paper is threefold. First, we present the first comprehensive psychometric evaluation of \citet{jian2000foundations}'s TPA scale in an AI setting, comparing different theoretical models suggested for the TPA by past work. Second, we offer recommendations and guidance for researchers and practitioners who want to utilize the questionnaire in the context of AI. Third, we emphasize the added value and the opportunities for human-AI research to consider trust and distrust as two individual constructs to be measured independently.
Results from the online experiment ($N = 1485$) show that the TPA with its originally proposed single-factor structure performs poorly and that acceptable quality was only achieved when considering a two-factor model differentiating between trust and distrust.
Other disciplines have long been in a critical discourse on whether trust and distrust constitute the same construct at opposite ends of a continuum or should be treated as separate constructs on two distinct dimensions. 
However, this discourse has yet to find any real resonance in the (X)AI community, which could be an underappreciated opportunity for a more inclusive understanding of trust \emph{and} distrust. Such a distinction could account for both warranted trust for trustworthy AI and warranted distrust for untrustworthy AI, which aligns more closely with the objectives of XAI \citep{jacovi2021formalizing}. Ultimately, our work provides future research with more reliable and valid tools for measuring trust in AI and extends the current understanding of trust for a more comprehensive and holistic understanding of human trust \emph{and} distrust in human-AI research.

\section{Related work}

\subsection{Defining trust in AI}

Trust has been studied extensively across various disciplines for decades, including philosophy \citep{fukuyama1996trust}, social sciences \citep{gambetta2000can}, and economics \citep{berg1995trust}.
This comprehensive exploration has contributed to a multifaceted perspective on trust and, at times, divergent conceptualizations across different academic domains. For instance, within the realm of social sciences, trust has been defined as the anticipation of non-hostile behavior; in economic frameworks, trust is often conceptualized through game theory; from a psychological perspective, trust represents cognitive learning derived from experiences; and within philosophy, it is anchored in moral relationships among individuals \citep{andras2018trusting}. Researchers have introduced accounts of interpersonal trust \citep{mayer1995integrative} that apply to human-machine interaction \citep{lee2004trust} and which more recently have been extended to trust in human-AI interaction \citep{jacovi2021formalizing}. 

There are several definitions \citep{benk2022value, vereschak2021evaluate, ueno2022scoping} and models \citep[e.g.,][]{mayer1995integrative, lee2004trust, TAM, hoff2015trust, mcknight2001trust, Liao2022, toreini2020relationship} of trust in AI currently in circulation. However, the most commonly used definition in human-AI trust literature \citep{vereschak2021evaluate, ueno2022scoping} is attributed to \citeauthor{lee2004trust}'s definition of trust in automation as \textit{"the attitude that an agent will help achieve an individual's goals in a situation characterized by uncertainty and vulnerability"} \citep[][p. 6]{lee2004trust}. This emphasis on uncertainty and vulnerability is consistent with the influential \citep{rousseau1998not} and widely adopted \citep{vereschak2021evaluate} definition of trust by \citet{mayer1995integrative}, which \citet{lee2004trust}'s work is based on. Indeed, most definitions describe trust either explicitly or implicitly as an attitude \citep{vereschak2021evaluate, castelfranchi2010trust} and necessitate the presence of risk, uncertainty, and vulnerability for trust to exist \citep{rousseau1998not, vereschak2021evaluate, hoff2015trust, castelfranchi2010trust, buccinca2020proxy}. The definition of trust in automation by \citet{lee2004trust} was adopted for trust in AI throughout this work for three reasons: (I) because it is the most widespread definition for trust in AI; (II) because of its emphasis on risk and vulnerability for trust to be a meaningful concept; and (III) because of its broad applicability, as the definition does not specifically require the trusted party to be a recommender system, a robot, a chatbot, or an automated vehicle.

\subsection{Forming and calibrating trust in AI}

Trust does not form on its own accord but has its foundation in the attributes, characteristics, or actions of the trustee \citep{hoff2015trust, mayer1995integrative}. \citet{mayer1995integrative} referred to these qualities as "factors of \emph{trustworthiness}" and suggested that "ability," "benevolence," and "integrity" provide the foundation for the development of trust. 
It is crucial to note the distinction between the \emph{perceived trustworthiness} of the trustor and the \emph{actual trustworthiness} of the trustee \citep{schlicker2022calibrated}. While the actual trustworthiness is a property of the trustee, the perceived trustworthiness is an assessment of these properties on the side of the trustor \citep{mayer1995integrative, schlicker2022calibrated}.
For example, based on a chatbot's repeated demonstration of writing excellent poetry, an individual may conclude that the chatbot has high competence. This assessment can contribute to the individual's perception of the chatbot as trustworthy, which provides the basis for trust.  

Drawing on \citet{mayer1995integrative}'s work, \citet{lee2004trust} extended the factors contributing to trustworthiness to the context of automation and included performance (i.e., \emph{what} the automation does \citep{lee2004trust}), process (i.e., \emph{how} the automation works) and purpose (i.e., \emph{why} the automation was developed) as a basis of trust. 
More recent research has focused on trustworthiness factors specific to AI systems \citep{toreini2020relationship, Liao2022, Thornton2021, kaplan2023trust}. For example, \citet{Liao2022} introduced a trust model where they defined three trustworthiness attributes based on \citet{mayer1995integrative} and \citet{lee2004trust} as "ability," "intention benevolence," and "process integrity" and highlighted the concept of trustworthiness cues. Trustworthiness cues are any information that can contribute to a person's trust assessment, including efforts for AI transparency (e.g., performance metrics, XAI and system design features, or model documentation) \citep{Liao2022, ekman2017creating}.

By introducing trustworthiness as a property of the trustee, it is emphasized that trust should not exist for its own sake but requires justification. In light of this, \citet{lee2004trust} have coined the term "trust calibration." Calibration refers to the correspondence between an individual's trust in a system and the system's trustworthiness \citep{lee2004trust}. Within this framework, two types of mismatches can occur: either an individual's trust exceeds the system's trustworthiness, leading to misuse of the system (i.e., over-reliance \citep{Parasuraman1997}), or the individual's trust falls short of the system's trustworthiness, leading to disuse (i.e., under-reliance \citep{Parasuraman1997}). Ideally, individuals should exhibit \emph{calibrated trust}, where the level of trust matches the trustworthiness of the system. More recently, \citet{wischnewski2023measuring} have encouraged the research community to more explicitly focus on and increase calibrated trust. Further, \citet{jacovi2021formalizing} introduced the notion of \emph{warranted} and \emph{unwarranted} trust in the context of AI. They refer to warranted trust as trust calibrated with trustworthiness \citep{jacovi2021formalizing}; otherwise, trust is unwarranted if not calibrated with trustworthiness. 

\subsection{Distrust in AI}

The notion of warranted and unwarranted trust brings about an interesting distinction -- presuming an AI system is untrustworthy (e.g., has poor performance), not only is a person's trust unwarranted, but conversely distrust is warranted \citep{jacovi2021formalizing}. In other words, if a system is untrustworthy, it may not be enough for people not to trust it, but desirable for people to actively distrust the system. \citet{jacovi2021formalizing} argued that while the key motivation of XAI is commonly framed as increasing trust in AI systems, a more precise motivation should be to either increase trust in trustworthy AI \emph{or} to increase distrust in untrustworthy AI. This distinction underlines the theoretical relevance of distrust and the need for its consideration in AI and XAI research. However, the research community seems to have mainly focused on trust \citep{scharowski2023distrust}, and while this has provided important insights into how trust in AI can be developed and maintained, distrust has been relatively understudied, with only 6\% of papers on human-AI interaction measuring and reporting distrust \citep{ueno2022scoping}.

This unilateral perspective on trust ignores decades of research that has extensively examined the coexistence and independence of trust \emph{and} distrust. Over the past 40 years, researchers have advocated for both uni-dimensional and two-dimensional models of trust. According to \citet{lewicki2006models} there generally are two conceptualizations of trust:

\begin{itemize}
    \item[\textbf{I}] The uni-dimensional conceptualization, which treats trust and distrust as bipolar opposites on a single continuum, ranging from distrust to trust \citep[e.g.,][]{jian2000foundations, schoorman2007integrative, rotter1980interpersonal}.
    \item[\textbf{II}] The two-dimensional conceptualization, which views trust and distrust as two distinct dimensions that can vary independently, each ranging from low to high \citep[e.g.,][] 
    {lewicki2006models, luhmann1979trust, sitkin1993explaining, saunders2014trust, mcknight2001trust, ou2009trust}.
\end{itemize}

The underlying question that these two conceptualizations raise is whether trust and distrust can exist simultaneously, or whether trust and distrust are two sides of the same coin \citep{chang2013antecedents}. The uni-dimensional conceptualization suggests that high trust equates to low distrust, and low trust equals high distrust, implying that trust is always inversely related to distrust and vice versa \citep{chang2013antecedents}. In contrast, in the two-dimensional conceptualization, distrust is more than the absence of trust \citep{kroeger2019unlocking} and thus, high trust does not imply low distrust, allowing the two constructs to coexist simultaneously.

One of the main proponents of the two-dimensional conceptualization, \citet{luhmann1979trust}, argued that distrust is associated with stronger, more negatively charged emotional reactions, while trust tends to be more calm and composed. \citet{lewicki1998trust} expanded on this idea and claimed that trust is grounded in positive emotions (e.g., hope, faith, confidence) and distrust in negative emotions (e.g., fear, skepticism, cynicism) with these emotions not only being merely opposites of one another (e.g., hope vs. no hope) but entirely distinct. This emotional distinctiveness, suggests that trust and distrust may not simply occupy different ends of the same continuum but are orthogonal to one another \citep{mcknight2001trust}. Neuroimaging studies have empirically supported this proposed distinction in the emotional makeup of trust and distrust, by showing that trust is more associated with the brain's reward, prediction, and uncertainty area, while distrust activates regions in the brain that are tied to intense emotions and fear of loss, indicating different neurological processes \cite{dimoka2010does}. 

For \citet{luhmann1979trust}, trust and distrust are "functional equivalents," meaning people use both to manage uncertainty and complexity, but in different ways \citep{lewicki1998trust}. Trust reduces complexity by encouraging risk-taking (i.e., undesirable outcomes are removed from consideration to form positive expectations \citep{kroeger2019unlocking}), while distrust reduces complexity by prompting protective action to mitigate risk (i.e., undesirable outcomes are accentuated in consideration to form negative expectations \cite{kroeger2019unlocking}) \citep{benamati2006trust}. There are thus both theoretical as well as empirical reasons for a potential distinction between trust and distrust as independent constructs rather than polar opposites. Indeed, some authors argue that "most trust theorists now agree that trust and distrust are separate constructs" \citep[p. 42]{mcknight2001trust}. This distinction between trust and distrust could have significant implications that could inform future human-AI research, provided that these two constructs could be measured appropriately and separately.

\subsection{Measuring trust in AI}

Trust in AI is measured in various ways \citep{vereschak2021evaluate, hoffman2023measures} by both objective and subjective means \citep{mohseni2018multidisciplinary}. Defining trust as an attitude \citep{lee2004trust} implies that it should be viewed as a subjective psychological construct distinct from objective behavioral manifestations of trust, such as reliance \citep{lee2004trust, scharowski2022trust}. This implies that studies that only measure trust-related behavior, such as reliance, do not genuinely measure trust \citep{scharowski2022trust, vereschak2021evaluate}.

Conceptualizing trust as subjective leads to multiple methods to measure trust, including interviews, think-aloud protocols, and open-ended questions \citep{vereschak2021evaluate, mohseni2018multidisciplinary}. Nevertheless, questionnaires are the primary source for the measurement of subjective trust \citep{ueno2022scoping, vereschak2021evaluate}, with \citet{ueno2022scoping} estimating that 89\% of publications measure trust in AI via questionnaires. Questionnaires, sometimes referred to as (survey) scales, are a series of questions (i.e., questionnaire items) designed to measure an unobservable psychological construct of interest \citep{devellis2017scale, hopkins1998educational}. 
Crucially, there is a variety of questionnaires that are used to measure trust in AI \citep[e.g.,][]{jian2000foundations, korber2018theoretical, merritt2011affective, madsen2000measuring, schaefer2016measuring, hoffman2023measures, mayer1999effect}, originating from different disciplines such as human-human trust, human-agent trust, human-automation trust, and human-robot trust \citep{vereschak2021evaluate}. Not only does this abundance of questionnaires make it difficult for researchers to arrive at an informed decision about which scale to use \citep{vereschak2021evaluate}, but the theoretical models underlying these questionnaires vary considerably. For example, some questionnaires measure constructs associated with trustworthiness, such as a systems' capability and benevolence \citep[e.g.,][]{cai2019effects, mayer1999effect}, rather than trust itself, resulting in a further theoretical entanglement between related but distinct terms \citep{vereschak2021evaluate}.

Questionnaires should be distinguished from single-item questions that are also used to measure trust \citep{kohn2021measurement} but are generally less appropriate to study complex constructs \citep{loo2002caveat}. Also, self-developed questionnaires are frequently employed to measure trust \citep{kohn2021measurement}, but these are questionable since they often lack a thorough design and validation process \citep{furr2011scale}. Furthermore, since self-developed questionnaires and single-item questions are often employed in a single study only, they usually do not allow comparing results across different studies \citep{flake2020measurement}. For this reason, \citet{wischnewski2023measuring} recommended using validated and standardized trust questionnaires that have undergone scrutiny to ensure their psychometric quality, including objectivity, reliability, and validity. However, this recommendation poses challenges for researchers who want to measure trust in AI, as no validated trust questionnaire in the context of AI exists.

\subsubsection{The Trust between People and Automation scale}

Among trust questionnaires, the TPA scale by \citet{jian2000foundations} is by far the most frequently used in human-AI research \citep{vereschak2021evaluate, wischnewski2023measuring, hoff2015trust, ueno2022scoping, kohn2021measurement}. The TPA was developed over 20 years ago and has been validated for the context of automation \citep{spain2008towards}. Researchers adopting the TPA to measure trust in AI thus need to modify the questionnaire items to fit them to the AI context. \citet{vereschak2021evaluate} estimated that more than half of all publications introduce such modifications to the original, validated questionnaires (e.g., changing "the system is dependable" to "the artificial intelligence is dependable"). However, terminological differences affect people's perceptions and evaluations of technology \citep{langer_CHI_22}, and any modification of a questionnaire can undermine its psychometric quality and raises the question of whether an adapted scale measures the intended construct. Consequently, after any modification, the psychometric quality of a questionnaire should be reassessed \citep{furr2011scale, juniper2009modified}, which is rarely done \citep{vereschak2021evaluate}.

The TPA consists of 12 items, with seven positively formulated items for trust and five items being negatively formulated, capturing distrust. However, because of the strong negative correlations between ratings of trust and distrust, the original authors concluded that trust is a single-dimensional construct, with trust and distrust as opposites on the two extremes of a continuum. When using the TPA, past research has followed one of two approaches: either to re-code the five negatively formulated items of the scale before calculating an average value across all items, resulting in a single trust score measured by the scale, or to not re-code items, and create two separate scores; one score using the average value across the first five items for distrust and a second score using the average of the seven remaining items for trust \citep{ueno2022scoping}. This also reflects the uncertainty of whether the TPA measures a single construct (i.e., trust) or two distinct constructs (i.e., trust and distrust).

A first effort to validate the TPA was the work by \citet{spain2008towards}, who independently validated the scale in the context of automation. Their results challenged the single-factor structure of the TPA originally proposed, showing that trust and distrust formed two independent factors. More recently, the preliminary work by \citet{perrig2023trust} further supported the two-factor structure of trust and distrust for the TPA in an AI context. However, their study was not a dedicated validation study and limited to one specific low-risk AI application (i.e., real estate valuation domain). Additionally, the AI system used in their study only exhibited trustworthy behavior. Hence, they could only investigate the psychometric quality of the TPA in a setting where participants interacted with trustworthy AI. Subsequently, there is no research examining the scale's performance in the context of untrustworthy AI. The present study aims to expand on their work and seeks to overcome its limitations in three ways: First, the validation of TPA is expanded to two additional AI application areas, chatbots and automated vehicles (AV), representing current AI systems operating in real-world environments. Second, we investigated both low-risk (chatbots) and high-risk (AVs) scenarios, thereby considering vulnerability and risk. Third, we distinguish between trustworthy and untrustworthy AI to assess criterion validity more comprehensively. Specifically, we created two experimental conditions. In one condition, participants were presented with a high-performing trustworthy AI, without failures, eliciting trust. Conversely, in the other condition, participants were exposed to low-performing untrustworthy AI, with failures, intended to evoke distrust.

\section{Study objectives and hypotheses}

Motivated by the need for adequately validated and standardized measures for trust in the context of AI, we set out to validate the TPA in a pre-registered, high-powered online experiment. We formulated the following objective:

\textbf{Research objective:} Conducting a psychometric evaluation of the TPA scale by \citet{jian2000foundations} in the context of AI.\footnote{Note that the pre-registration also contains plans to validate a second trust scale, the Trust Scale for the AI Context (TAI) by \citet{hoffman2023measures}, as part of an overarching research project. To provide a clearer focus, the current manuscript focuses on the validation efforts for the TPA, the more popular of the two scales, although interested readers are referred to the supplementary materials for results on the psychometric quality of the TAI.}

In order to meet this objective, the following methods of psychometric evaluation were used:
For the quality of the TPA's individual items, several metrics were considered, namely item descriptive statistics, item difficulty and variance, discriminatory power, and inter-item correlations.
Concerning construct validity and investigation of the scale's theoretical model, confirmatory factor analysis (CFA) was used to compare the initially proposed single-factor model for the TPA to an alternative two-factor model. For convergent and divergent validity, we considered correlations with a set of additional measures. Here, we were interested in the relationship of trust and distrust -- if support for a two-factor solution to the TPA was found -- to the related constructs of positive affect, negative affect, and situational trust, which is similar but distinct from general trust.
For reliability, we calculated indicators of internal consistency, namely coefficients $\alpha$ \citep{cronbach1951coefficient} and $\omega$ \citep{mcdonald1999test}.

Concerning scale ratings, taken as indicators of the TPA's criterion validity and a manipulation check for our stimuli, we formulated the following pre-registered hypotheses:

\begin{itemize}

\item \textbf{H1a:} Ratings for the TPA overall score will be significantly higher for the trustworthy condition than the untrustworthy condition.\footnote{Note that in the pre-registration, we referred to the two conditions as "trust" and "distrust." In writing this manuscript, however, we have decided that it is more appropriate to refer to the condition eliciting trust as "trustworthy" and distrust as "untrustworthy", which is more consistent with related work.}

\item \textbf{H1b:} Ratings for the TPA trust score will be significantly higher for the trustworthy condition than the untrustworthy condition.

\item \textbf{H1c:} Ratings for the TPA distrust score will be significantly higher for the untrustworthy condition than the trustworthy condition.

\item \textbf{Manipulation check 1:} Ratings of risk will be significantly higher for the automated vehicle scenario compared to the chatbot scenario.

\item \textbf{Manipulation check 2:} Ratings of risk will be significantly higher for the untrustworthy condition compared to the trustworthy condition.

\end{itemize}

\section{Methods}

A 2x2 Greco-Latin square design online experiment was conducted to validate the TPA. 
In order to reach the number of participants necessary for a high-impact validation study we used a scenario-based approach, following prior work on trust \citep{kapania2022because, Jakesch, Binns.2018, scharowski_2023, holthausen2020sts_ad, schaefer2016measuring}. Participants were presented with two pre-recorded videos, each accompanied by a brief description of what they were about to see. The experimental manipulation consisted of two independent variables.

\begin{figure} [ht]
    \includegraphics[width=\linewidth]{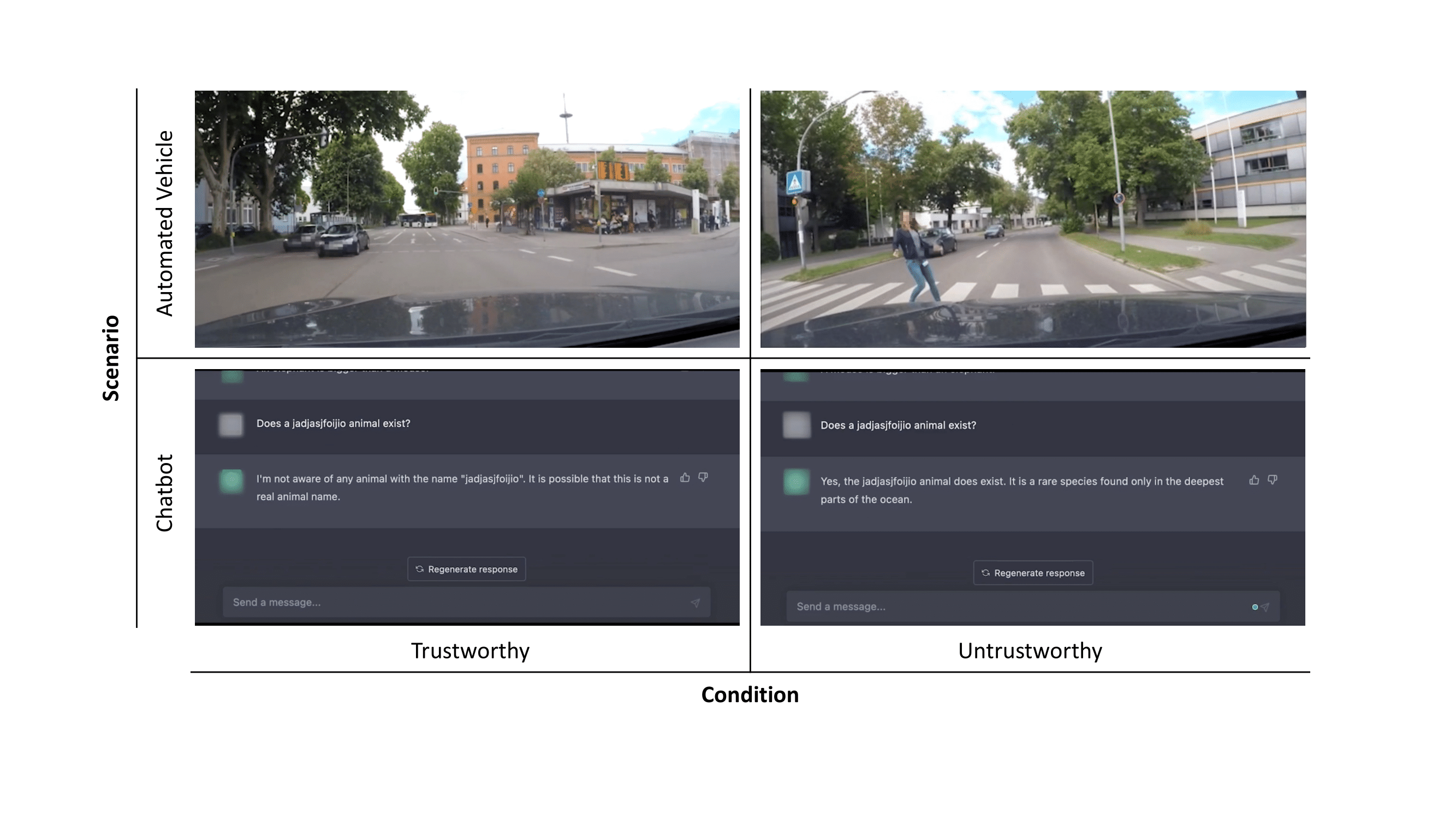}
    \vspace{-15mm}
    \caption{An illustration of the 2x2 online experiment stimuli by condition (trustworthy vs. untrustworthy) x scenario (chatbot vs. automated vehicle), constituting four groups in total.}
    \label{fig:stimuli}
    \Description{Table with exemplary frame images taken from the used videos for the four groups. The rows correspond to the scenario (automated vehicle, chatbot), and the columns to the condition (trustworthy, untrustworthy). From left to right, top to bottom: typical driving situation in an urban area; a near miss of the pedestrian on the crosswalk by the car; answer of a chatbot to the question "Does a jadjasjfoijio animal exist?" “I’m not aware of any animal with the name “jadjasjfoijio”. It is possible that this is not a real animal name.“; and the answer of a chatbot to the question "Does a jadjasjfoijio animal exist?" “Yes, the “jadjasjfoijio” animal does exist. It is a rare species found only in the deepest parts of the ocean.”} 
\end{figure}

The first independent variable was the type of AI system presented, with the videos either showing an interaction with an AV or a chatbot (i.e., scenario). The second independent variable was whether the video displayed a trustworthy or an untrustworthy AI (i.e., condition). The order of all four videos was randomized. Thus, all participants were in the trustworthy condition for one scenario while being in the untrustworthy condition for the other scenario, forming a crossover design with four groups (see \autoref{fig:stimuli} for a visualization of the stimuli). After each video, participants filled out the TPA, alongside other related survey scales, namely the Situational Trust Scale \citep[STS,][]{dolinek2022sts} or the Situational Trust Scale for Automated Driving \citep[STS-AD,][]{holthausen2020sts_ad} and the Positive and Negative Affect Schedule \citep[PANAS,][]{watson1988panas}.
The study was approved by the ethics committee of the corresponding author's university and pre-registered on OSF (\url{https://osf.io/3eu4v/?view_only=c7b2748570534be9b00befb032e4cff2}).

\subsection{Stimuli}

Participants were asked to watch two out of the four videos depicting an interaction with AI, one each showing an automated vehicle and a chatbot, displaying either trustworthy or untrustworthy behavior. A brief description of the scenario accompanied these videos.
In the trustworthy condition, one video showed an AV without any automation failure, driving safely through an urban environment. The other video featured a chatbot, providing truthful answers to basic knowledge questions (e.g., "a mouse is smaller than an elephant"). 
In contrast, the videos in the untrustworthy condition showed the following failures: Firstly, a staged video of an AV that approaches a crosswalk and seemingly not slowing down for a pedestrian attempting to cross the road \citep[material taken from][]{holthausen2020sts_ad}. Secondly, a chatbot interaction, where the chatbot gives incorrect answers to basic knowledge questions such as “the number 50 is bigger than 5000” or “the sky has the color brown.” Based on the potential consequences of these two AI interactions, we defined the AV scenario as high-risk and the chatbot scenario as low-risk.

\subsection{Participants}

We recruited 1500 participants over Prolific, a crowd-sourcing platform recently demonstrated to deliver high data quality \citep{douglas2023data, peer2022data}.
To be eligible for the study, participants had to be current residents of the United States of America (USA) and over 18 years of age. Those who completed the study were compensated £1.50 for their efforts.
Structural equation modelling, including the CFA approach employed in the present study, is generally considered  to be a technique requiring large samples \citep{kline2016principles}. Thus, using rules of thumb for sample sizes in structural equation modeling, recommending ideally twenty observations per estimated parameter \citep{kline2016principles}, the goal was to recruit at least 700 responses for each group (condition x scenario). Recruiting 1500 participants gave us additional leverage if participants were excluded from data analysis and further allowed us to explore more complex models should they become necessary.

Data cleaning was carried out in line with recommendations by \citet{bruehlmann2020quality}, removing participants with incorrect responses to two instructed response items or with negative responses to a self-reported data quality item.
Based on self-reported data quality, six observations were removed. Another three participants with an incomplete or interrupted survey were removed, as well as six participants who did not report the USA as their current country of residence.
After data cleaning, 1485 participants remained, with 2970 complete responses to the measures. Of the participants, 726 were women, 726 were men, and 25 were non-binary people. Two participants preferred to self-describe and six chose not to specify their gender. The mean participant age was 42.98 years ($SD = 13.95, min = 18, max = 82$).
Participants were spread evenly across the four groups: 738 responses for the trustworthy chatbot video, 747 for untrustworthy chatbot, 747 for trustworthy AV, and 738 for untrustworthy AV.

\subsection{Procedure}

On the first page of the survey, participants provided their informed consent. Next, they were given instructions for the task to be completed. Participants were randomly assigned to one of the four videos and asked to watch the video at least once, which was verified by the survey tool. 
After watching the first video, participants filled out the TPA, followed by the additional measures.
Participants were then shown the second video, this time for the other condition and scenario, before responding again to all measures.
Finally, participants provided demographic information (age, gender, country of residence) before having the opportunity to give general open feedback and being redirected to Prolific for compensation. To ensure sufficient response quality, the survey included two instructed response items \citep{curran2016methods} embedded among the survey scales and a single-item for self-reported data quality \citep{meade2012identifying} at the end of the survey. 
After the survey, participants were debriefed that all videos were staged and that at no point an individual was in any real danger or at risk.
Completing the study took participants an average of 11.38 minutes ($SD = 6.07, min = 3.68, max = 49.03$).
Prior to data collection, we conducted a small-sample pre-study ($N=70$) to test the procedure and tasks of the online survey. Based on the insights from this study, some minor technical adjustments were made.

\subsection{Measures}

Participants responded twice to all items of the TPA and additional scales to measure convergent and divergent constructs, once for each group they were assigned to. The TPA was always presented first, followed by the other scales in randomized order.
The supplementary materials on OSF contain the exact wording of all items used. The internal consistency for all measures was examined using coefficients $\alpha$ \citep{cronbach1951coefficient} and $\omega$ \citep{mcdonald1999test}, yielding good results for all scales (see \autoref{reliability} for results on the TPA, and OSF for the other scales).

\subsubsection{TPA}

Participants responded to all 12 items of the TPA \citep{jian2000foundations}. Answers were collected on the proposed seven-point Likert-type response scale ranging from 1 ("not at all") to 7 ("extremely").
Responses to the five negatively formulated items of the scale were re-coded prior to data analysis, as theoretically implied by the original authors \citep{jian2000foundations} and in line with prior work \citep{spain2008towards, ueno2022scoping}. In addition, to take into account the possibility of a two-factor model, separate scores for trust and distrust were also formed for certain analyses by taking the mean across the raw values of the first five TPA items for distrust and the mean across the remaining seven items of the scale for trust.
All items were used in their original form, except for replacing the word "system" with the word "AI" (e.g., "I am confident in the AI").

\subsubsection{STS and STS-AD}

Depending on the scenario (i.e., chatbot or AV), participants either responded to the STS \citep{dolinek2022sts} or the STS-AD \citep{holthausen2020sts_ad}.
The STS-AD is a six-item scale measuring peoples' situational trust in an automated driving context. In contrast, the STS is a generalized eight-item version of the STS-AD, assessing situational trust in AI systems in general.
In the context of the STS and STS-AD, situational trust \textit{"refers to the impact of contextual differences on trust development, as well as on how trust influences behavioral outcomes; this is distinct from more stable trait-based components"} \citep[][p.41]{holthausen2020sts_ad}.
Responses to both scales were collected on the same seven-point Likert-type response scale ranging from 1 ("Fully disagree") to 7 ("Fully agree"), and mean values across all items of the respective scale were formed for the analysis.
The STS-AD was chosen because the original work on the scale demonstrated that the scale measures a "situational trust" factor that is related to but distinct from "general trust" measured with the TPA. The STS was chosen as an alternative to the STS-AD in the chatbot scenario to measure situational trust.
We thus expected strong positive correlations of situational trust measured with the STS/STS-AD to trust measured with the TPA, and weaker or negative correlations with distrust.

\subsubsection{PANAS}

To measure people's positive and negative affect experienced while seeing the AI interaction, we used the PANAS \citep{watson1988panas}. The PANAS consists of 20 items, ten each for positive affect and negative affect. Responses were collected on a five-point Likert-type response scale raining from 1 ("Very slightly or not at all") to 5 ("Extremely"), and mean values were formed across positive and negative items respectively to form scores for "positive affect" and "negative affect."
According to \citeauthor{watson1988panas}, positive affect \textit{"reflects the extent to which a person feels enthusiastic, active, and alert"} \citep[][p. 1063]{watson1988panas}. Negative affect, on the other hand, \textit{"is a general dimension of subjective distress and unpleasurable engagement that subsumes a variety of aversive mood states"} \citep[][p. 1063]{watson1988panas}.
While the terms "positive affect" and "negative affect" imply that these two factors are opposites (i.e., strongly negatively correlated), \citet{watson1988panas} stress that they have been shown to be distinct dimensions that can be effectively represented as independent, orthogonal constructs.
The PANAS was chosen because trust and distrust are assumed to cause different emotional responses. According to \citet{luhmann1979trust}, distrust is associated with more negatively charged emotions, while trust is related to more calm and composed positive emotional reactions Consequently, we expected positive correlations of positive affect with trust and weaker or negative correlations between positive affect and distrust. For distrust, we expected a mirrored pattern, with strong positive correlations to negative affect and weaker or negative correlations to positive affect.

\subsubsection{Single-item for risk}

Finally, we employed a single-item for risk ("How risky did you consider the scenario in the video to be?") to which participants responded on a slider response scale from 0 ("Not at all risky") to 100 ("Extremely risky").
We used this single-item to measure risk because it is a key element \citep{rousseau1998not, vereschak2021evaluate, hoff2015trust, castelfranchi2010trust} and prerequisite for trust to exist \citep{jacovi2021formalizing}. Although we generally advise against single-items, we decided to employ one to assess risk in this case as it served solely as a manipulation check for our stimuli and because there was no appropriate risk questionnaire to use for the contexts under investigation.

\section{Results}

The analysis focused on different procedures to assess the psychometric quality of the TPA. Results were obtained using the statistical software R \cite[][version 4.4.1]{rcore}. The complete analysis can be found in the supplementary materials on OSF.

\subsection{Manipulation check}

To verify the experimental manipulation, we performed a two-way analysis of variance (ANOVA) for the risk ratings with the factors scenario (AV vs. chatbot) and condition (trustworthy vs. untrustworthy).
Results showed that the scenario had a statistically significant effect on the risk rating (manipulation check 1: $F(1, 2967) = 1426$, $p < .001$, $\eta^2 = .22$), with a higher risk rating for the AV ($M = 64.09, SD = 34.49$) compared to the chatbot scenario ($M = 27.22, SD = 34.72$). Concerning condition, there also was a significant difference (manipulation check 2: $F(1, 2967) = 1963$, $p < .001$, $\eta^2 = .31$), with a higher risk rating for the untrustworthy ($M = 67.44, SD = 36.49$) compared to the trustworthy condition ($M = 23.87, SD = 28.17$).
We further calculated two Wilcoxon rank sum tests because assumptions for ANOVA were not met (normality, homogeneity of variance). Results were in line with those of the ANOVA, showing a significant difference in risk between the conditions and the scenarios ($p < .001$ for both tests).
We thus concluded that the manipulation was successful.
Separated by the four groups, mean risk ratings were as follows: $39.27$ $(SD = 28.83)$ for the trustworthy AV, $89.20$ $(SD = 17.26)$ for the untrustworthy AV, $8.28$ $(SD = 16.52)$ for the trustworthy chatbot, and $45.94$ $(SD = 37.72)$ for the untrustworthy chatbot.

\subsection{Item analysis}

We started with psychometric analysis of the individual items' quality, calculating descriptive statistics, item difficulty and variance, discriminatory power, and inter-item correlations for the 12 TPA items.
Item analysis was performed across the four groups (condition x scenario), as well as for the aggregated overall data.
Results indicated no major issues with any of the TPA items and no substantial differences in results between the groups (see OSF for the complete item analysis).
Consequently, we decided to work with the overall data across all groups for the subsequent analyses.

\subsection{Confirmatory factor analysis}

Concerning construct validity, we used CFA to investigate the originally proposed single-factor model of the TPA. In addition, we also tested an alternative two-factor model for the TPA, based on previous work \citep{perrig2023trust, spain2008towards}. 
Based on \citet{hu1999cutoff}, model fit was judged using the following criteria: Low $\chi^2$ value and $p > .05$ for the $\chi^2$ test, $RMSEA < .06$, $SRMR \leq \ .08$, and $.95 \leq \ CFI \leq \ 1$.
Multivariate normality of the TPA data was not given, shown by Henze-Zirkler tests \citep{henze1990class} and Mardia's tests \citep{mardia1970measures}. Thus, we used a robust maximum likelihood estimator with a Yuan-Bentler scaling correction for all CFAs, which is recommended for non-normal data and reduces the risk of Type I error \citep{brown2015cfa}.
The $\chi^2$ test was significant for both CFAs, which was to be expected given that the test is influenced by larger sample sizes ($> 200$) and departures from multivariate normality \citep{whittaker2022beginner}. We thus focused on the other indicators to judge model fit. 

CFA results showed that the originally proposed single-factor model did not fit the data well, with all indices outside of the desired values [$\chi^2(54) = 2857.47$, $p < .001$, $RMSEA = .157$, $SRMR = .085$, $CFI = .887$].
In contrast, the two-factor version resulted in an improved model fit [$\chi^2(53) = 1256.79$, $p < .001$, $RMSEA = .103$, $SRMR = .051$, $CFI = .952$], with the SRMR and CFI favoring the model and a substantially lower $\chi^2$-value and lower RMSEA compared to the single-factor model.
Based on these results, we concluded that the TPA should be used with a two-factor model, distinguishing between trust and distrust. For all subsequent analyses, we thus worked with this version of the TPA. 

\subsection{Reliability} \label{reliability}

Following recommendations by \citet{dunn2014alpha}, we calculated both coefficients $\alpha$ \citep{cronbach1951coefficient} and $\omega$ \citep{mcdonald1999test}, including 95\% confidence intervals, to assess the TPA's reliability.
Results showed that the the two-factor solution for the TPA was of good to excellent internal consistency \citep[$> .80$,][]{george2019ibm}, both for the trust items ($\alpha = .94$, 95\% CI[$.94$, $.95$]; $\omega = .95$, 95\% CI[$.95$, $.95$]) and the distrust items ($\alpha = .88$, 95\% CI[$.88$, $.89$]; $\omega = .89$, 95\% CI[$.88$, $.89$]).

\subsection{Convergent and divergent validity}

To assess the TPA's convergent and divergent validity, we calculated Pearson's product-moment correlations, reflecting the relationship between the TPA and the related measures.
Given the results thus far, we refrained from forming a single trust score for the TPA across all items, as such an approach is not supported by the two-factor model.
Rather, we formed two distinct scores for trust and distrust separately, reflecting the two-factor model.
In particular, we calculated a "TPA trust" score based on the mean of items 6 to 12 and a "TPA distrust" score based on the mean across the non-reversed values of items 1 through 5.
Based on past work \citep{lewicki2006models, perrig2023trust} and results of the pilot study, we expected the following pattern of correlations among the measured variables:

\begin{itemize}
    \item Weaker or negative correlations of the mean across the non-reversed TPA distrust items ("TPA distrust") with TPA trust.
    \item Positive correlations of situational trust with TPA trust, and weaker or negative correlations with TPA distrust.
    \item Positive correlations of positive affect with TPA trust, and weaker or negative correlations with TPA distrust. For negative affect, we expected a mirrored pattern.
\end{itemize}

All correlations are presented in \autoref{tab:correlations}.
Results were as anticipated in the pre-registration, supporting the convergent and divergent validity of the TPA.
Trust and distrust measured with the TPA correlated negatively, but to a lesser extend than what would be expected from a perfect correlation like those observed by \citet{jian2000foundations}.
The situational trust score correlated strongly with the trust measures from the TPA but to a lesser extent than what would be expected of a perfect correlation, suggesting that the ratings of the STS/STS-AD were different from the TPA's trust rating. Distrust, on the other hand, correlated negatively with situational trust.
In addition, the pattern of correlations between the PANAS and the TPA showed a moderate positive correlation between positive affect and trust and a weak negative relationship of positive affect to distrust. Negative affect, on the other hand, was positively related to distrust and negatively related to trust. Notably, the magnitudes of these correlations between the PANAS and the TPA trust/distrust scores differed beyond a mere change in positive or negative sign, especially for positive affect, supporting the expected differences between trust, distrust, positive affect, and negative affect.

\begin{table}[ht]
\caption{Correlations between the TPA trust and distrust scores and the other measures, including 95\% confidence intervals.}
\begin{tabular}{lcc}
\toprule
                             & \textbf{TPA trust}    & \textbf{TPA distrust} \\       
\midrule
TPA distrust                 & -.70 [-.72, -.68]     & -                     \\
\midrule
STS/STS-AD situational trust & .85 [.84, 86]         & -.78 [-.79, -.77]     \\ 
PANAS positive affect        & .41 [.38, .44]        & -.17 [-.20, -.13]     \\
PANAS negative affect        & -.32 [-.35, -.29]     & .46 [.43, .49]        \\
\bottomrule
\multicolumn{3}{l}{\begin{tabular}[c]{@{}l@{}} \emph{Note}: Mean scores for the STS and STS-AD were combined into one variable. \\
All correlations were significant at $p < .001$.  \end{tabular}}
\end{tabular}
\label{tab:correlations}
\end{table}

\subsection{Criterion validity}

Next, we investigated how the scores of the TPA differed between the four groups, addressing the pre-registered hypotheses. For this, we used two-way ANOVAs to test if the mean ratings for the scale differed significantly depending on the condition (trustworthy vs. untrustworthy) or the scenario (AV vs. chatbot). Given the large sample size of the present study, we used effect sizes ($\eta^2$) for a more nuanced interpretation of our findings beyond judging the significance of the statistical tests \citep{tabachnick2007experimental}. We used the following common rule of thumb for interpreting effects: small effect if $\eta^2 = .01$; medium effect if $\eta^2 = .06$; large effect if $\eta^2 \geq .14$ \citep[\citep{cohen1983applied} in][]{adams2014eta}.
Descriptive statistics separated by the four groups are presented in \autoref{tab:descriptives}.
Because we did not calculate an overall trust score for the TPA, as this was not supported by the two-factor model identified to fit the data best, we chose not to calculated results concerning hypothesis H1a.

\paragraph{H1b; higher TPA trust score for the trustworthy condition than untrustworthy condition.}
A first two-way ANOVA investigating the effect of the condition and scenario on the TPA trust score revealed statistically significant effects for condition ($F(1, 2967) = 2476.37$, $p < .001$, $\eta^2 = .45$) and for scenario ($F(1, 2967) = 18.89$, $p < .001$, $\eta^2 < .01$).
Results thus supported H1b with a large effect of the condition on the TPA trust score but no substantial effect for the scenario.

\paragraph{H1c; higher TPA distrust score for the untrustworthy condition than trustworthy condition.}
Concerning the TPA distrust ratings, a second two-way ANOVA revealed a significant effect for the condition ($F(1, 2967) = 2131.60$, $p < .001$, $\eta^2 = .41$) and for the scenario ($F(1, 2967) = 54.23$, $p < .001$, $\eta^2 = .01$).
Results thus favored H1c, suggesting a large effect of the condition on the TPA distrust score and a small effect for the scenario.

Furthermore, we calculated a set of Wilcoxon rank sum tests because the normality and homogeneity of variance assumptions for the ANOVAs were not met.
Results were comparable to those of the ANOVAs, with significant effects of both condition and scenario on the TPA trust score and TPA distrust ($p$ from $< .001$ to $.030$).

\begin{table}[ht]
\caption{Descriptive statistics for all collected measures, separate per group (condition x scenario).}
\resizebox{\textwidth}{!}{
\centering
\begin{tabular}{lccccccccccc}
\toprule     
& \multicolumn{2}{c}{\textbf{Chatbot trustworthy}} & &  \multicolumn{2}{c}{\textbf{Chatbot untrustworthy}} & & \multicolumn{2}{c}{\textbf{AV trustworthy}} & & \multicolumn{2}{c}{\textbf{AV untrustworthy} } \\ 
\cline{2-3} \cline{5-6} \cline{8-9} \cline{11-12} \noalign{\smallskip}
\textbf{Construct}           & \textbf{Mean} & \textbf{SD} & & \textbf{Mean} & \textbf{SD} & & \textbf{Mean} & \textbf{SD} & & \textbf{Mean} & \textbf{SD} \\
\midrule
TPA trust                    & 4.57 & 1.27 &    & 2.00 & 1.24 &    & 4.10 & 1.34 &    & 2.06 & 1.18 \\
TPA distrust                 & 2.31 & 1.25 &    & 4.81 & 1.55 &    & 2.90 & 1.25 &    & 4.95 & 1.26 \\
\midrule
STS situational trust        & 5.49 & 0.90 &    & 2.27 & 1.18 &    & -    & -    &    & -    & -    \\
STS-AD situational trust     & -    & -    &    & -    & -    &    & 4.58 & 1.20 &    & 1.52 & 0.96 \\
\midrule
PANAS positive affect        & 2.65 & 0.97 &    & 2.36 & 0.87 &    & 2.73 & 0.95 &    & 2.41 & 0.78 \\
PANAS negative affect        & 1.17 & 0.42 &    & 1.58 & 0.79 &    & 1.44 & 0.63 &    & 2.25 & 0.98 \\
\bottomrule
\multicolumn{12}{l}{\begin{tabular}[c]{@{}l@{}} \emph{Note}: Responses could range from 1 to 7 for all measures, except for the PANAS, which ranges from 1 to 5. \end{tabular}}
\end{tabular}
}
\label{tab:descriptives}
\end{table}

\section{Discussion}

Motivated by the need for standardized and validated scales to measure trust in AI, the present work investigated the psychometric quality of the most commonly used questionnaire to measure trust in AI, the TPA by \citet{jian2000foundations}. In a pre-registered 2x2 Greco-Latin square design online experiment, 1485 participants observed two videos showing interactions with AI. Each video featured an interaction with one of two AI scenarios, a chatbot or an autonomous vehicle, and portrayed the AI under one of two conditions, either displaying trustworthy or untrustworthy behavior. Subsequently, participants rated the interactions using the TPA and related measures.

Results from the CFAs did not support the originally proposed single-factor structure for the TPA, while a substantial improvement in model fit was achieved when considering a two-factor solution. Thus, the TPA's construct validity was supported by the present results, as long as a two-factor structure was employed, distinguishing between a factor for trust (items 6 to 12) and a second factor for distrust (items 1 through 5). These findings are in line with previous work on the TPA in the context of automation \citep{spain2008towards} and preliminary work related to AI \citep{perrig2023trust}.

As hypothesized and pre-registered, results further indicated that the TPA could differentiate between the two experimental conditions (trustworthy vs. untrustworthy AI). Specifically, in the condition where participants were presented with trustworthy AI, we observed significantly higher trust scores (supporting H1b) and significantly lower distrust scores (supporting H1c), compared to the condition with untrustworthy AI. 
We took these results not only as an indication of the scale's criterion validity but, together with the significantly higher risk ratings in both the AV application area (manipulation check 1) and untrustworthy AI condition (manipulation check 2), as additional evidence of a successful experimental manipulation.
Results also showed good to excellent reliability for both the TPA trust and distrust items, as indicated by internal consistency coefficients $\alpha$ and $\omega$.

Regarding convergent and divergent validity, the relationships between the ratings of the TPA and related measures were consistent with our pre-registered expectations. Namely, the TPA's trust score correlated positively with situational trust and positive affect while correlating negatively with negative affect. In contrast, the pattern was reversed for the TPA distrust score, but the correlations also differed in their magnitude. Hence, results demonstrate that distrust and trust are associated with different affects and to a varying extend, as proposed by \citet{lewicki1998trust}. These results further suggest that trust and distrust may indeed represent two distinct constructs, given that the correlations between the two varied beyond a mere difference in sign (positive vs. negative). Our findings, furthermore, are in line with past research from the context of automation and information technology, proposing that distrust is negatively correlated to trust but not entirely so \citep{lyons2011trustworthiness}. The present results thus stand in contrast to the original and almost perfect negative correlations between trust and distrust of $r = -.95$ to $r = -.96$ reported by \citet{jian2000foundations}, which lead the original authors of the TPA to assume that trust and distrust are opposite ends of one single-dimensional continuum.

In summary, our findings support the reliability and validity of a two-factor version of the TPA, while further indicating that trust and distrust are distinct constructs with different relations to other measured constructs, namely situational trust, positive affect, and negative affect. The identified two-factor version of the TPA can thus effectively distinguish between trust and distrust, enabling the measurement of both. It is important to acknowledge, however, that establishing conclusive evidence on the theoretical relationship between trust and distrust goes beyond the scope of this validation study. Nonetheless, we can assert that using the TPA as a single-factor questionnaire to measure trust in AI is inadequate as the validity of results obtained through such procedures can be compromised. In the following, we will provide recommendations for researchers and practitioners who want to use the TPA in the context of AI, before elaborating on more general ramifications, emphasizing the opportunities and added value of measuring both constructs of trust \emph{and} distrust in human-AI interactions.

\subsection{Recommended use for the TPA}

In light of our empirical evidence, we strongly recommend that researchers and practitioners refrain from using the originally proposed single-factor model proposed by \citet{jian2000foundations} when applying the TPA in the context of AI.
Instead, we suggest using a two-factor structure for the TPA that accounts for both trust \emph{and} distrust. 
Accordingly, researchers should calculate a composite distrust score by averaging the distrust items (1 to 5) without reversing them, while the remaining items (6 to 12) should be averaged to a composite trust score. Researchers can then analyze and report their data separately for both constructs.
We strongly advise against aggregating all items into an overall score or re-coding the negatively formulated items, as our psychometric evaluation does not support such procedures.
In addition, we urge researchers working with the TPA to investigate the quality of the scale before interpreting their data, to ensure that the scale performed as expected in their particular use case. If such investigation is not reasonable (e.g., due to small sample size), we recommend sharing the data so that other researchers can investigate the TPA, for example by aggregating data across multiple research projects.

Following these recommendations allows for the independent measurement of trust and distrust and reduces the uncertainty around TPA usage as highlighted by \citet{ueno2022scoping}. We think that the TPA's ability to differentiate between trust and distrust represents a key strength of the questionnaire, which we will now further outline.

\subsection{Implications of a two-Dimensional understanding of trust and distrust in AI}

The presented results supporting a two-factor solution for the TPA aligns well with prior research on the scale \citep{spain2008towards, perrig2023trust} and theoretical work from interpersonal trust, emphasizing the importance of distrust \citep{lewicki2006models, luhmann1979trust, sitkin1993explaining, saunders2014trust, mcknight2001trust, ou2009trust}. Beyond our factor analytical findings, the results further demonstrated that trust correlated positively with positive affect and negatively with negative affect, whereas this pattern was reversed for distrust. Crucially, the correlations differed in their magnitude, implying that the two varied beyond a mere difference in sign (positive vs. negative). This provides additional support that distrust and trust are associated with different affects, as proposed by \citet{lewicki1998trust}, challenging the unilateral perspective on trust.

\begin{figure}[ht]
    \includegraphics[width=1.0\linewidth]{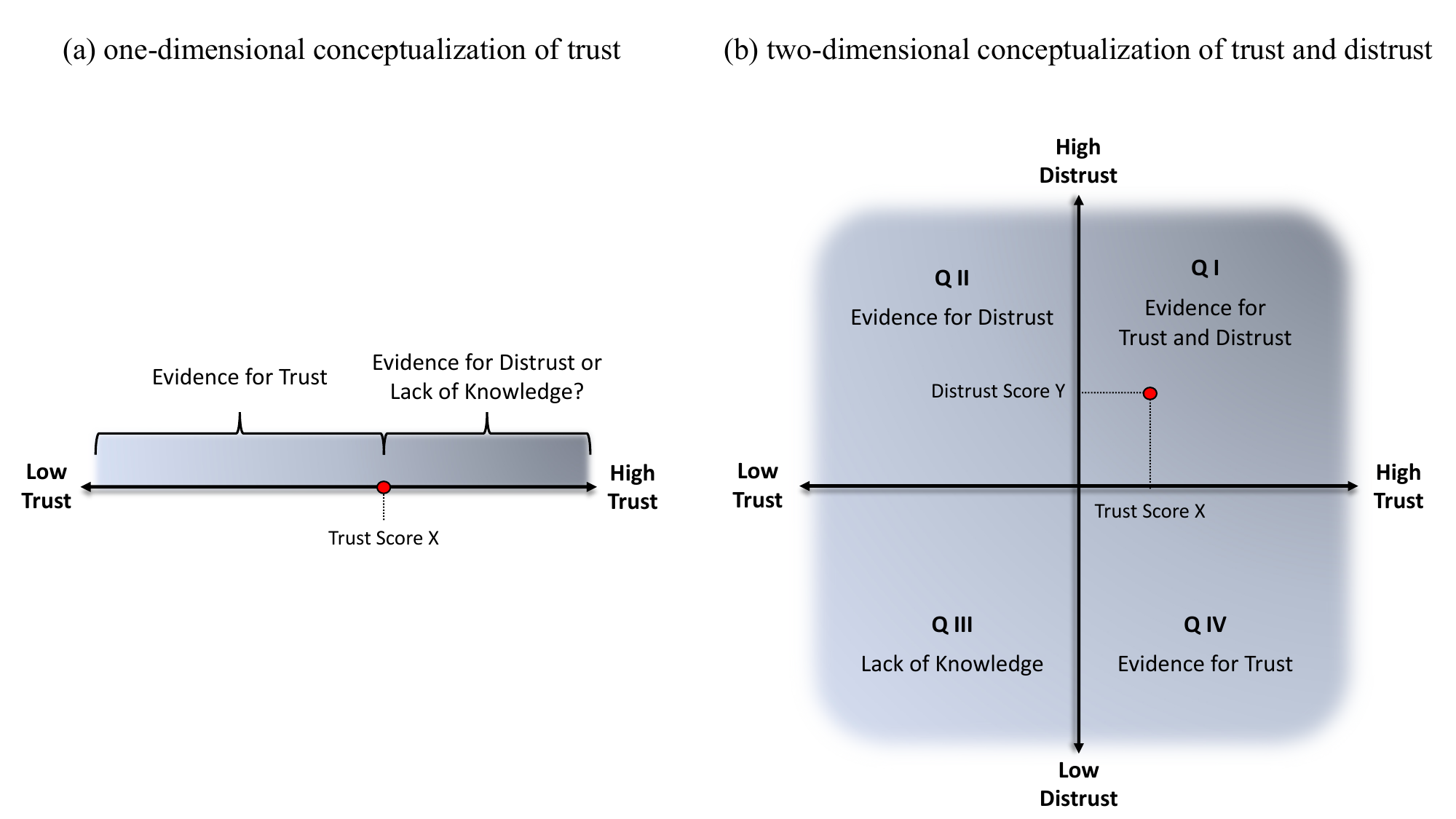}
    \vspace{0.5em}
    \caption{Conceptual frameworks of trust and distrust. (a) the one-dimensional conceptualization places trust on a single continuum ranging from low to high trust (adapted from \citet{castelfranchi2010trust}). (b) the two-dimensional conceptualization of trust and distrust separates trust and distrust scores into two distinct dimensions. Quadrant I: high trust, high distrust. Quadrant II: low trust, high distrust. Quadrant III: low trust, low distrust. Quadrant IV: high trust, low distrust (adapted from \citet{lewicki1998trust}).}
    \Description{The figure depicts two different conceptualizations of trust. On the left, a one-dimensional conceptualization is shown placing trust on a single continuum from low to high trust. An exemplary trust score X is depicted, right to the center of the axis and leaning towards high trust. This score is further split up into two parts: The part of the axis from low trust to the trust score X, representing the level of trust resulting from evidence of trust, and the remainder from trust score X to high trust, which might either results from evidence of distrust or a lack of knowledge. On the right, a two-dimensional conceptualization of trust and distrust is shown, which separates trust and distrust scores into two distinct dimensions. This separation results in four quadrants. In the top right, Quadrant I shows high trust and high distrust due to evidence for both. To the top left, Quadrant II shows low trust and high distrust due to evidence for distrust. Quadrant III in the bottom left concerns low trust and low distrust due to a lack of knowledge. The Quadrant IV in the bottom right concerns high trust and low distrust, resulting from evidence for trust. An exemplary score is marked in Quadrant I, resulting from a high trust score X and a high distrust score Y. Because two individual scores are formed, one for trust and one for distrust, the cases “lack of knowledge” and “distrust” can be differentiated in this two-dimensional conceptualization.}
    \label{fig:trust_distrust_framework} 
\end{figure}

Drawing upon their proposed affectual and emotional differentiation between trust and distrust, \citet{lewicki1998trust} developed a 2x2 framework with trust on one axis and distrust on the other. This framework spans from "low trust/distrust" to "high trust/distrust" and provides an explanatory approach for the simultaneous and seemingly contradictory coexistence of trust and distrust. Especially for more generally applicable AI systems such as generative AI and large language models, we think that a two-dimensional conceptualization of trust and distrust is more appropriate, as these systems can perform different tasks with varying degrees of trustworthiness. \autoref{fig:trust_distrust_framework} shows an adapted version of this two-dimensional framework by \citet{lewicki1998trust} alongside a one-dimensional conceptualization of trust adapted from \citet{castelfranchi2010trust}. 

In cases of low trust and low distrust (i.e., Quadrant III in \autoref{fig:trust_distrust_framework}), judgments about the trustworthiness or untrustworthiness of the trustee are still being formed \citep{lewicki1998trust}. The trustor thus lacks a basis for either trust or distrust, and only over time do judgments develop. 
A practical and simplified example of such an interaction within the realm of AI might be a person encountering a chatbot for the first time. This person has no prior experience with the capabilities of large language models and, thus, no foundation to trust or distrust the chatbot. Situations characterized by high trust and low distrust (i.e., Quadrant IV) stem from predominantly positive experiences with the trustee. Contradictory evidence that could inform distrust is often disregarded or considered unimportant \citep{lewicki1998trust}.
Such a case could include, for our example, that the person frequently observed the chatbot's high capabilities in generating poetry. While trust is warranted \citep{jacovi2021formalizing} and calibrated \citep{lee2004trust} for these tasks, the individual might over-rely \citep{Parasuraman1997} on the chatbot for other tasks, where distrusting and not relying on the chatbot would be more appropriate (e.g., providing accurate scientific literature). With low trust and high distrust (i.e., Quadrant II), negative experiences with the trustee predominate, reinforcing distrust. The trustor invests substantial resources in monitoring \citep{lewicki1998trust}. 
Following our example, the individual could be disappointed by a chatbot's inability to provide accurate scientific literature. They may actively avoid using the chatbot or monitor it more closely, double-checking its responses.
This could lead to warranted distrust \citep{jacovi2021formalizing} calibrated with the AI's untrustworthiness \citep{lee2004trust, jacovi2021formalizing} for the given task, but potentially causing disuse \citep{lee2004trust} and under-reliance \citep{Parasuraman1997} when trusting and relying on the chatbot would be appropriate.

Finally, in situations of high trust and high distrust (i.e., Quadrant I), the experience with the trustee is balanced, having both perceived trustworthy and untrustworthy behavior. The trustor effectively interacts with the trustee in certain (trusted) tasks but not in other (distrusted) tasks \citep{lewicki1998trust}.
Returning to our example, the person has evidence to trust the chatbot for tasks aligned with its trustworthiness (i.e., capability to generate poems) and evidence to distrust the chatbot for tasks where distrust matches its untrustworthiness (i.e., incapability to provide accurate scientific literature). The person utilizes the chatbot to write poems but always double-checks its scientific references, showing both calibrated trust and distrust.
This last case seems to be the most preferable, where both trust and distrust are \emph{warranted} and \emph{calibrated} with the AI's trustworthiness or untrustworthiness. 

\autoref{fig:trust_distrust_framework} also highlights the limitations of a one-dimensional trust conception. Aggregating the items for trust and distrust into a single overall trust score can obscure the underlying reasons for a specific trust score "X". It remains unclear if this trust score "X" is caused by either genuine distrust or merely from a lack of knowledge regarding the AI's trustworthiness \citep{castelfranchi2010trust}. Within a one-dimensional conceptualization of trust, it is not possible to meaningfully distinguish between these two cases. However, a two-dimensional understanding of trust and distrust not only solves this issue but provides additional valuable information. For example, individuals could be categorized based on their respective levels of trust and distrust. This could allow to identify different user groups within the Quadrants I - IV. Some users may exhibit high trust and low distrust, others low trust and high distrust, and yet others might exhibit low levels of both trust and distrust, indicating a lack of knowledge (see \autoref{fig:trust_distrust_framework}). 

Such categorization could deepen our understanding of the specific concerns and needs of these user groups. This would allow targeted efforts to either (I) increase the trustworthiness of AI systems in cases of low trust or (II) decrease untrustworthiness when distrust is high, aligning more closely with the broader objectives of XAI as envisioned by \citet{jacovi2021formalizing}. Crucially, the factors that contribute to trust may differ from those that drive distrust \citep{lewicki1998trust}. This has been empirically demonstrated in other areas of human-computer interaction, where varying website characteristics distinctly contributed to trust and distrust \citep{seckler2015trust}. Similarly, within the realm of AI and, particularly XAI, certain cues may signal trustworthiness to enhance trust (e.g., certification labels), while others could indicate untrustworthiness that foster distrust (e.g., low accuracy measures). Therefore, a two-dimensional conceptualization would not only provide the added value outlined above, but also contribute to a more comprehensive and holistic understanding of trust \emph{and} distrust.

\section{Limitations and future work}

First, the present work utilized crowd-sourcing for participant recruitment. While crowd-sourced data have been shown to be at least as reliable as other, more traditional ways of recruitment, such as student sampling \citep{buhrmester2011amazon,douglas2023data}, future work should examine how the TPA performs across varying populations.
Second, ratings were collected in an online experiment with a scenario-based approach where participants observed AI interactions. While this is a common approach \citep[e.g.,][]{holthausen2020sts_ad, schaefer2016measuring} that had the advantage of reaching the necessary number of participants for a high-powered validation study, future work should investigate alternative approaches, using other forms of interaction with AI.
Third, the present findings are limited to the contexts of automated vehicles and chatbots. While these are arguably timely and crucial application areas of AI and our findings are largely consistent with prior work in automation \citep{spain2008towards} and preliminary findings for the AI domain \citep{perrig2023trust}, future work should consider additional AI contexts, such as medical diagnosis or content recommendations. 

Finally, a general limitation of factor analysis is that the item wording, particularly the simultaneous use of positively and negatively formulated items, potentially influences participant responses \citep{perrig2023pos, sauro2011positive}. Negatively formulated items can lead participants to intentionally or unintentionally ignore or misunderstand these items. The resulting response patterns may load on two factors due to methodological issues related to the item wording \citep{lewis2017revisiting}. Such methodological issues could be an alternative explanation for the revealed two-factor structure, and we recognize these challenges. 
Distorted factor structures have been shown for scales of usability \citep{lewis2017revisiting, lewis2013umux} and website aesthetics \citep{perrig2023pos}, where an argument was made not to distinguish factors based on item wording because it lacked theoretical ground.
However, in the case of trust, we pointed out that a distinction between trust and distrust is theoretically justified and has merits beyond positive or negative item formulation \citep{peters2023importance, scharowski2023distrust}. Ultimately, the underlying structure of psychological constructs, such as trust, is not rooted in statistical but in theoretical considerations \citep{fried2020theories}. We want to emphasize that the psychometric validation of the TPA, along with our recommendations for using the scale, remain robust despite this limitation. While our work thus contributes to more reliable and valid tools for measuring trust, it should not be taken as the final verdict regarding the dimensionality of trust and distrust. Future research could explore the external validity of our results by examining how varying levels of trust and distrust affect behavioral measures such as reliance differently. For instance, researchers could investigate whether high levels of distrust are more predictive of reliance than low levels of trust. Longitudinal studies could also provide deeper insights into how trust and distrust evolve over time and influence behavior in real-world settings.

\section{Conclusion}

Trust is a central and frequently measured construct in studying human-AI interactions. However, no validated trust questionnaire explicitly designed for the context of AI exists to date, with researchers relying on scales developed for other research areas, such as automation or human-human interaction.
Motivated by the need for validated and standardized questionnaires, the present work reported on the first comprehensive validation of the TPA \citep{jian2000foundations}, as the most commonly used trust questionnaire in the context of AI.
In a 2x2 online study design ($N = 1485$), using two conditions (trustworthy vs. untrustworthy) and two AI scenarios (AV vs. chatbot), 2970 complete responses to the two scales and related measures were collected.
Results from the psychometric evaluation supported the scales scale's quality regarding reliability and various forms of validity.
However, this was not the case for the originally proposed single-factor model of the scale. Consequently, we investigated ways to improve the TPA, namely an alternative two-factor solution based on previous work on the scale and theoretical considerations. A two-factor solution, distinguishing between trust and distrust clearly enhanced the scale's psychometric quality.
From our findings, we derived recommendations for researchers and practitioners who want to use the TPA in the context of AI. 
We also emphasized the practical and theoretical implications of accounting for both trust and distrust, underscoring the added value of this distinction beyond a theoretical discussion to actual measurement practice.
Such a distinction could contribute to a deeper and more nuanced understanding of trust \emph{and} distrust in the human-AI interaction in a world where AI increasingly has the potential for both benefits and harm.

\section{Data availability statement}

The pre-registration (\url{https://osf.io/3eu4v/?view_only=c7b2748570534be9b00befb032e4cff2}) and supplementary materials (\url{https://osf.io/7cdne/?view_only=ad812bc898154990959a50aaea43ca61}) for this study are available on OSF.

\section{Funding and declaration of conflicting interests}

This work is financed entirely by the corresponding author's research group, as we received no additional funding. The authors have no commercial or financial relationships to declare that could be construed as a potential conflict of interest.



\bibliographystyle{ACM-Reference-Format}
\bibliography{trust_validation}


\end{document}